\documentclass{iaus}
\usepackage{graphicx}

\title[The RAVE harvest] 
{The RAVE harvest:\\from the relation between abundances and kinematic of the
Milky Way stars to tools for the abundance analysis of the spectra}

\author[Corrado Boeche]   
{Corrado Boeche$^1$
 \and the RAVE collaboration
}

\affiliation{$^1$Astronomisches Rechen-Institut\\ Zentrum f\"ur Astronomie
der Universit\"at Heidelberg\\ M\"onchhofstr.
12-14, 69120 Heidelberg, Germany\\ email: {\tt
corrado@ari.uni-heidelberg.de}}
\pubyear{2013}
\volume{xxx}  
\pagerange{1--6}
\jname{Setting the scene for Gaia and LAMOST}
\editors{A.C. Editor, B.D. Editor \& C.E. Editor, eds.}
\begin{document}

\maketitle

\begin{abstract} RAVE is a spectroscopic survey of the Milky Way which
collected more than 500,000 stellar spectra of nearby stars in the Galaxy.  
The RAVE consortium analysed
these spectra to obtain radial velocities, stellar parameters and chemical
abundances.  These data, together with spatial and kinematic information like positions,
proper motions, and distance estimations, make the RAVE database a rich source
for galactic archaeology.  I present recent
investigations on the chemo-kinematic relations and chemical gradients in the
Milky Way disk by using RAVE data and compare our results with the Besan\c con
models.  I also present the code SPACE, an evolution of the RAVE
chemical pipeline, which integrates the measurements of stellar parameters
and chemical abundances in one single process.
\end{abstract}

\firstsection 

\section{The RAVE survey} 

The RAdial Velocity Experiment (RAVE) is a large spectroscopic survey which
observed stars of the Milky Way (Steinmetz et al.,
\cite{steinmetz2006}) in the magnitude interval 9$<I<$12
in the southern celestial hemisphere.  After ten years of observations, the
survey ended in April 2013, collecting 574,630 spectra of 483,330 stars. 
The spectra were obtained with the 1.2 meter UK Schmidt Telescope of
the Australian Astronomical Observatory.  The 150 optical fibres of the 6dF
spectrograph allowed us to collect up to 130 spectra in
one hour of exposure time.  With a resolution of R$\sim$7500, the spectra
centered on the near infrared Ca {\scshape ii} triplet region (8410-8795\AA)
yield precise radial velocity (RV) measurements ($\sigma_{RV}\sim$2~km~s$^{-1}$ 
at S/N$>$40). Beside, RAVE provides stellar parameters such as effective
temperature, gravity, and metallicity (Zwitter et al. 
\cite{zwitter2008}, Siebert et al. \cite{siebert2011}, Kordopatis et
al.  in preparation), and chemical abundances for the elements Mg, Al, Si,
Ti, Fe, and Ni (Boeche et al.  \cite{boeche2011}, Kordopatis et al.  in
preparation).
Proper motions of the RAVE stars come from a variety of catalogues, such as 
Tycho2 (H\o g et al., \cite{hog}), PPM-Extended catalogues PPMX and PPMXL 
(Roeser et al., \cite{roeser2008}, \cite{roeser2010}) and the second and 
third U.S. Naval Observatory CCD Astrograph Catalog UCAC2 and UCAC3 
(Zacharias et al., \cite{zacharias}).
Distances have been estimated with different methods (Breddels et al. 
\cite{breddels}, Zwitter et al.
\cite{zwitter2010}, Burnett et al. \cite{burnett}, Binney et al. in
preparation). The kinematic information, combined with the distances and the
chemical abundances, enables us to locate the RAVE stars in the six dimensional
phase-space and the chemical space, making the RAVE database a rich mine for
Galactic archaeology studies.\\
RAVE is complementary to the Sloan Extension for Galactic Understanding and
Exploration survey (SEGUE, Yanny et al. \cite{yanny}). 
In fact, the two surveys observe in opposite hemispheres, at different
magnitude ranges (SEGUE covers $14<g<20$), and at different
resolution (R$\sim$2000 for SEGUE). Because of the difference in magnitude, 
the SEGUE dwarf stars cover a volume similar to the one probed by the RAVE
giant stars, making the comparison of the two samples crucial for the
robustness of the results obtained with the two surveys.

\begin{figure}[t]
\begin{center}
 \includegraphics[width=6cm]{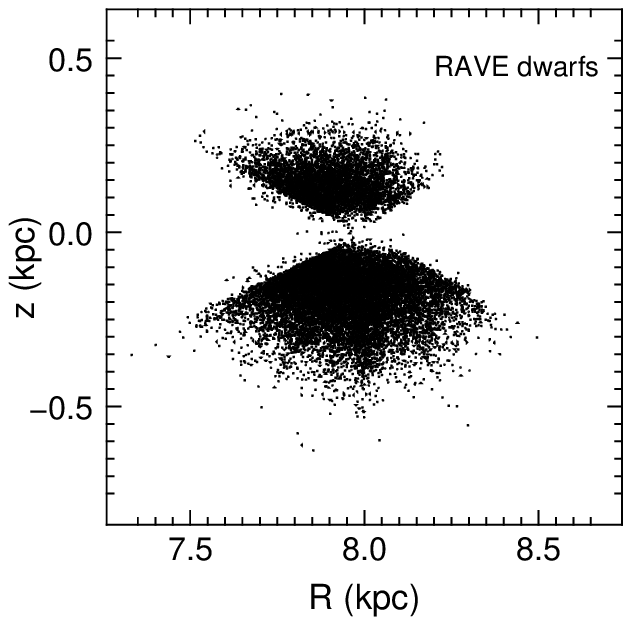} 
 \hspace*{0.5 cm}
 \includegraphics[width=6cm]{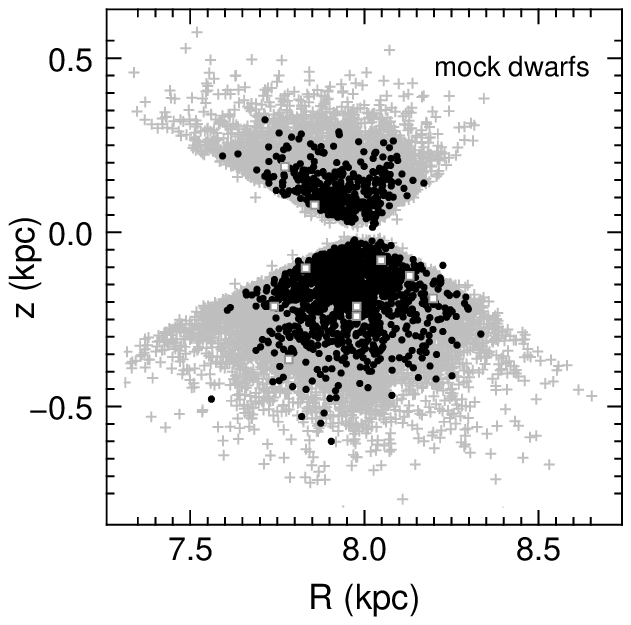} 
 \caption{Spatial distribution of the RAVE sample (left) and the mock sample
 (right) on the  meridional plane. For the mock sample the gray plus
 symbols, the black points and the gray open squares indicate the thin disc, the thick
 disc, and the halo stars, respectively.}
   \label{Boeche_Corrado_fig1}
\end{center}
\end{figure}

\section{Radial chemical gradients of the Galaxy with RAVE: a comparison
with the Besan\c con model}
\begin{figure}[t]
\begin{center}
 \includegraphics[width=12cm]{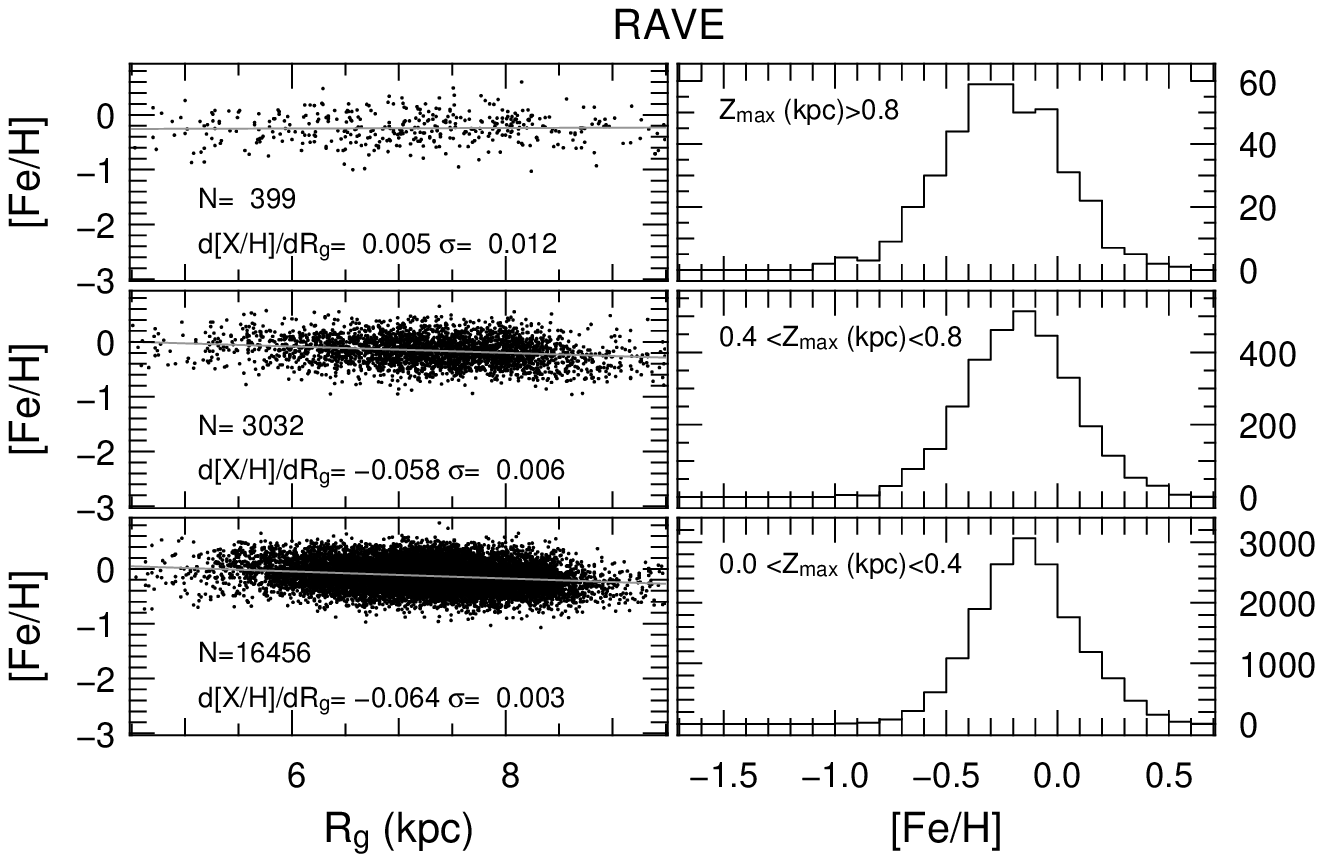} 
 \vspace*{0.5 cm}
 \includegraphics[width=12cm]{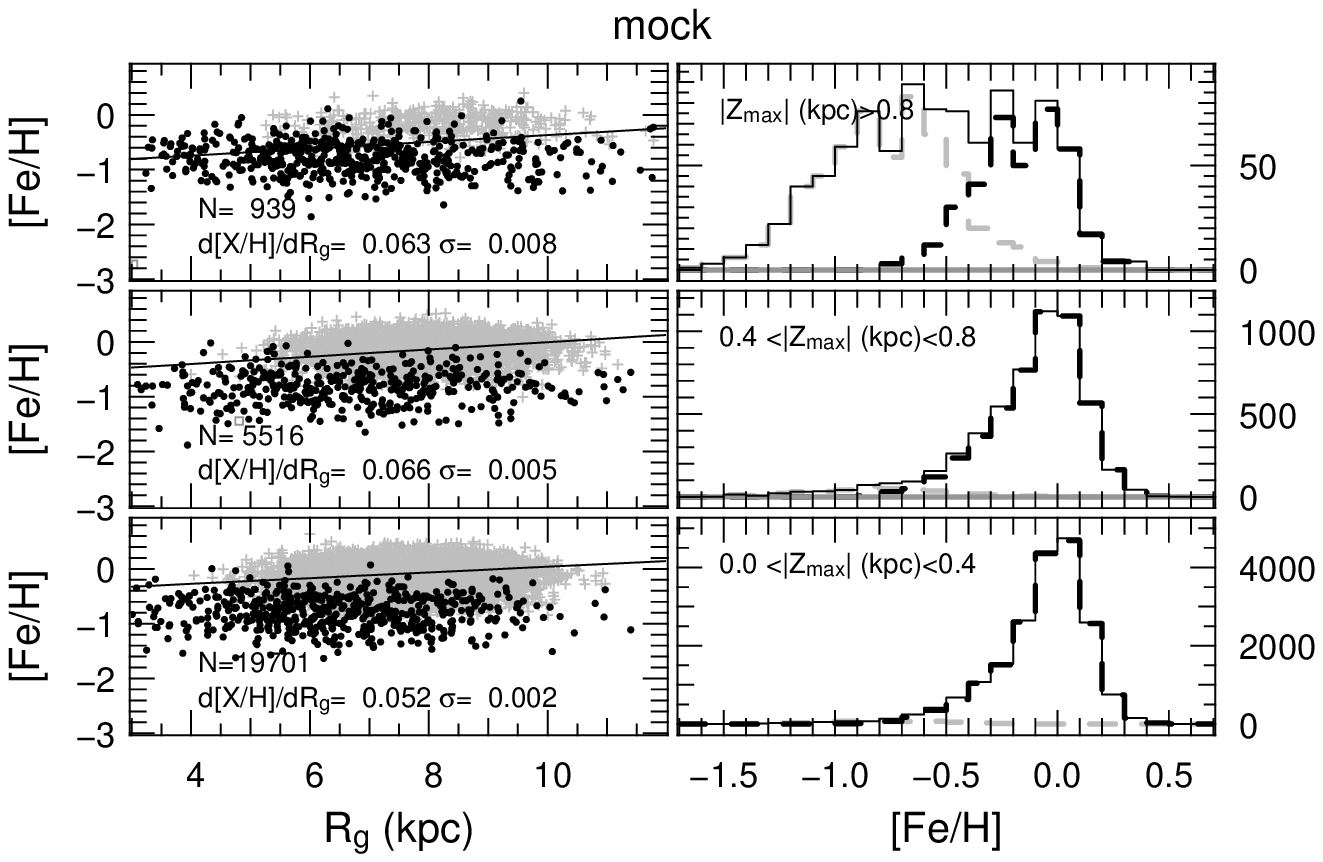} 
 \caption{Distribution of the stars in the ($R_g$,[Fe/H]) plane (left
 panels) and [Fe/H] distributions (right panels) of the three subsamples at
 different Z$_{\mbox{max}}$ for the RAVE sample (top) and the mock sample
 (bottom). Symbols are as in Fig.\,\ref{Boeche_Corrado_fig1}. 
The [Fe/H] distributions of the
 mock sample are traced with a dashed black line, a dashed gray line and a
solid grey line for the thin disc, thick disc and halo stars, respectively. The
 overall distributions are represented by a black thin line.}
   \label{Boeche_Corrado_fig3}
\end{center}
\end{figure}

We measured the chemical gradients of the Milky Way along its radius by
using a sample of 19,962 RAVE dwarf stars (Boeche et al., submitted)
selected to have spectra with S/N$>$40, effective temperature
5250$<T_{\mbox{eff}}(K)<7000$, gravity $\log g>$3.8~dex and error in distance
smaller than 30\%. The stars are also classified as normal stars by Matijevi{\v c} et al.
(\cite{matijevic}) and have little or no continuum defects in their spectra.
Since the RAVE dwarf stars cover a small Galacticentric distance range
($\sim$0.6~kpc, see Fig.\,\ref{Boeche_Corrado_fig1}, left panel) a chemical 
gradient measured by using the 
actual positions of the stars will be necessarily poorly constrained. Thus, we extended
the measurements to the Galactic radius range R$\sim$4.5--9.5~kpc by using
the guiding radius $R_g$. To obtain $R_g$, the Galactic orbits 
of the stars have been integrated in the potential
model n.2 by Dehnen \& Binney (\cite{dehnen}). From the rotation curve and the
integrated orbits we computed the guiding radius $R_g$, and extract other
orbital parameters such as apocentre, pericenter and maximum distance from
the Galactic plane reached by the star (Z$_{\mbox{max}}$).\\
In order to avoid observational bias and compare the real Galaxy with
models, we created a RAVE equivalent mock sample by using the stellar population
synthesis code GALAXIA (Sharma et al., \cite{sharma}), which uses
analytical density profiles based on the Besan\c con model (Robin et al.,
\cite{robin}). The mock sample reproduces the RAVE selection
funcion in I magnitude and the target distribution on the sky. The same cuts
in S/N, T$_{\mbox{eff}}$, $\log g$ applied to the RAVE
sample have been applied to the mock sample, which contains 26,198 entries (see
Fig.\,\ref{Boeche_Corrado_fig1}, right panel).
We measured the gradients of iron abundance [Fe/H] of these two samples at
three different Z$_{\mbox{max}}$ values by dividing them in three subsamples:
stars with $0.0<Z_{\mathrm{max}}$ (kpc)$\leq 0.4$, $0.4<Z_{\mathrm{max}}$ (kpc)$\leq 0.8$,
and $Z_{\mathrm{max}}$ (kpc)$>0.8$. The results are illustrated in
Fig.\,\ref{Boeche_Corrado_fig3} for the RAVE sample and the mock sample.

The RAVE sample exibits a negative gradient of
d[Fe/H]/d$R_g=-0.064\pm0.002$~dex~kpc$^{-1}$ for $Z_{\mathrm{max}}<0.4$~kpc, and
becomes flat at $Z_{\mathrm{max}}>0.8$~kpc. This result is in agreement with other
previous works (Cheng et al.  \cite{cheng}, Pasquali \& Perinotto
\cite{pasquali}).  Conversely, the mock sample shows positive gradients at
any $Z_{\mathrm{max}}$.  These unrealistic gradients have several causes: 
i) the mock sample has an excess of thick disc stars (see the [Fe/H]
distributions in Fig.\,\ref{Boeche_Corrado_fig3}, bottom plot, right
panels), ii) their mean metallicity appears too low with respect the RAVE
sample, and iii) because of the larger asymmetric drift and the lower metallicity of the
thick disc stars with respect to the thin disc stars, such stars are shifted
toward lower $R_g$ and lower [Fe/H] (Fig.\,\ref{Boeche_Corrado_fig3},
bottom plot, left panels). The superposition of thin and thick disc stars
mimics a positive gradient, the value of which depends on the ratio of thin/thick disc
stars in the sample. Besides, the thin disc stars of the mock sample have
the unrealistic gradient d[Fe/H]/d$R_g=0.00$~dex~kpc$^{-1}$ although
the gradient in the actual Galactocentric distance $R$ 
(assigned by the Besan\c con model) is d[Fe/H]/d$R=-0.07$~dex~kpc$^{-1}$. 
This difference is due to the absence of a correlation between the kinematics and
the metallicity. In fact, in the real Galaxy, stars with high eccentricity are more likely
to be metal poorer, whereas in the Besan\c con model the metallicities of the
stars are assigned considering their Galactocentric distances but
regardless of their eccentricities (i.e. kinematics). The discrepancies between the RAVE sample and
the mock sample can therefore be reduced by i) decreasing the density, ii) decreasing the
vertical velocity, iii) increasing the metallicity of the thick disc in
the Besan\c con model, and iv) assigning metallicities to the stars as a
function of their kinematics, so that stars in high eccentricity orbits are
on average metal poorer.

\begin{figure}[t]
\begin{center}
 \includegraphics[width=12cm]{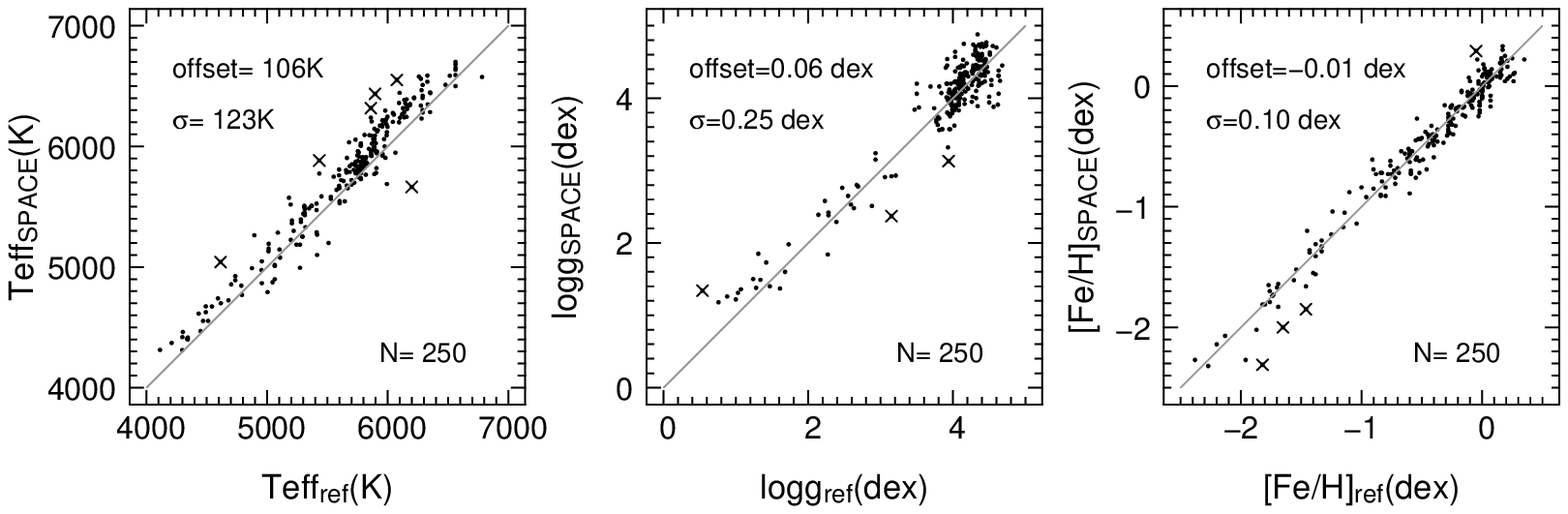} 
 \caption{Stellar parameters derived by SPACE (y-axis) and the ELODIE
reference parameter (x-axis) of ELODIE spectra reduced to spectral resolution
R=5000, S/N$\sim$70 and wavelength range 5200-6200\AA. We consider here only spectra having ``good" and
``excellent" parameters estimation flag in the ELODIE library.}
   \label{Boeche_Corrado_fig5}
 \vspace*{0.5 cm}
 \includegraphics[width=12cm]{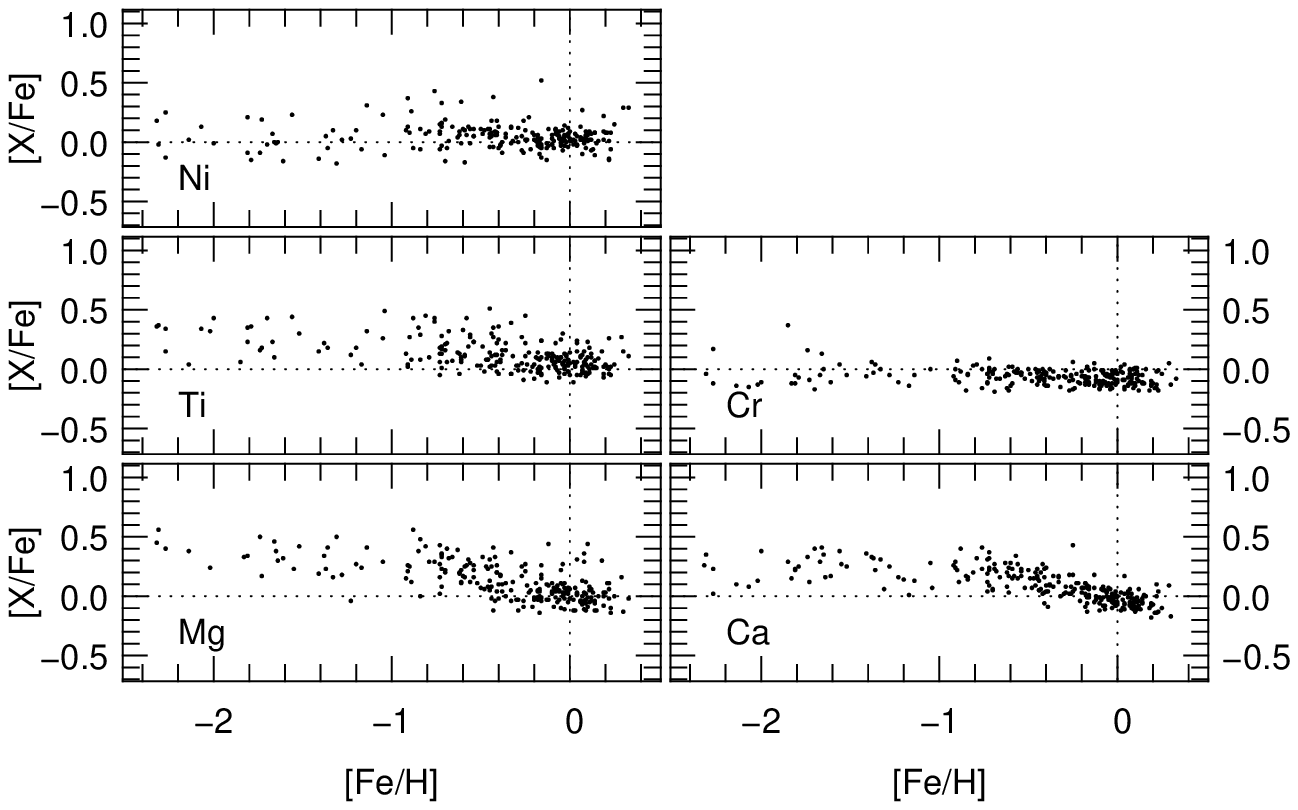} 
 \caption{Chemical abundances of the ELODIE spectra reduced to spectral
resolution R=5000, S/N$\sim$70 and wavelength range 5200-6200\AA
(the same used for Fig.\,\ref{Boeche_Corrado_fig5}).}
   \label{Boeche_Corrado_fig6}
\end{center}
\end{figure}

\section{SPACE: a new code for stellar parameters and chemical abundances
estimations}
Large spectroscopic surveys face the challenge to process and analyse large
databases of spectra in a reasonable time. This boosts the effort of
developing new methods and automated tools for spectral analysis, 
which are also part of the survey's outcome. 
SPACE (which stands for Stellar PArameters and Chemical
abundances Estimator) is one of RAVE's fruits, since it
evolved from the RAVE chemical pipeline (Boeche et al. \cite{boeche2011}). 
Both codes are based on 1D LTE (one dimensional, Local Thermodynamic Equilibrium)
atmosphere models. The RAVE chemical pipeline derives chemical abundances 
from a normalised, radial velocity corrected spectrum, where stellar parameters 
such as $T_{\mbox{eff}}$, $\log g$ and a first guess
of [M/H] are provided by an external source.
SPACE estimates stellar parameters and chemical abundances with no need of
extra information but the normalized spectra and a first guess of the
spectral resolution. 
SPACE does not rely on a library of synthetic spectra,
nor does it measure equivalent widths (EWs) of isolated lines. Instead it relies on a
list of lines with astrophisically corrected oscillator strenghts, and on a library of
Generalised Curves-Of-Growths (GCOGs) of such lines. The GCOG is a function
in the 3-dimensional parameter space (PS,
where the variables are $T_{\mbox{eff}}$, $\log g$, and abundance [X/H]), which
describes the variation of the EW of a line in the PS. When $T_{\mbox{eff}}$ and $\log g$ are
fixed, the GCOG reduces to the classical curve-of-growth.
SPACE retrieves the EWs of the
lines from the GCOGs of one point in the PS, reconstructs a spectrum model
by assuming a Gaussian/Voigt line profile and varies the stellar parameters
and chemical abundances, searching for the model that matches best the
observed spectrum via $\chi^2$ minimization. 
To date, SPACE works in the stellar parameter ranges of $4000<T_{\mbox{eff}}\mbox{(K)}<7000$, $0.0<\log
g\mbox{(dex)}<5.0$ and $-2.5<\mbox{Fe/H](dex)}<0.5$. Eextensions of the PS are
possible. In principle, SPACE can work 
in any wavelength range when an appropriate line list and the
corresponding library of GCOGs are provided. To date, SPACE works in the wavelength
ranges 5200-6200\AA\ and 8400-8900\AA. Other wavelength ranges will be
included soon.
We tested SPACE by using spectra of the ELODIE spectral
library (Prugniel et al., \cite{prugniel}) degraded to a
resolution of R=5000 and R=20\,000, a S/N$\sim$70, and considering only the wavelength
interval 5200-6200\AA. At resolution R=5000
(Fig.\,\ref{Boeche_Corrado_fig5}) SPACE gives satisfactory results with
$1\sigma$ errors in $T_{\mbox{eff}}$, $\log g$ and [Fe/H] of $\sim$120~K
(with an offset of +106~K), 0.25~dex and
0.10~dex, respectively. Chemical abundances have uncertainties
smaller than 0.1-0.2~dex (depending on the element) and correctly trace the
enhancement of $\alpha$-elements with respect to iron
(Fig.\,\ref{Boeche_Corrado_fig6}). At R=20\,000 the errors do not seem
smaller than the ones at R=5000 ($\sigma_{T_{eff}}$=172~K, $\sigma_{\log
g}$=0.33~dex, and $\sigma_{[Fe/H]}$=0.10~dex) because a not yet identified 
systematic error in $T_{\mbox{eff}}$ and $\log g$ affects the results.
Nonetheless, the resulting chemical abundances look unexpectedly good,
highlighting the gap in $\alpha$-enhancement between thin and thick disc
stars in [Ca/Fe] and [Ti/Fe]. More work and tests are needed in order to identify the causes of the systematic
errors and to extend the working wavelength range. Once a stable
version of the code exists, SPACE will be released to the scientific community as
a tool for spectral analysis.


\begin{discussion}

\discuss{Hou}{Can your SPACE software be used for spectra with different
resolution?}

\discuss{Boeche}{Yes, as showed in preliminary tests, SPACE gives reliable
results between resolution R=5000 and 20,000. I will test the code at
resolution R=2000 to see if surveys like SEGUE and LAMOST can profit from it.
Higher resolutions may be possible too and will be tested.}

\discuss{Ludwig}{How do you handle the microturbulence when you calculate
your library of curve-of-growths for SPACE?}

\discuss{Boeche}{The microturbulence is determined by a formula (given in
Boeche et al. \cite{boeche2011}) which is function
of T$_{\mbox{eff}}$ and $\log g$.}

\discuss{Ritter}{How do your stellar atmospheric parameters compare to the
official RAVE values?}

\discuss{Boeche}{I did not test SPACE on RAVE spectra yet, but I will do it
soon.}

\discuss{Monari}{In Fig.~\ref{Boeche_Corrado_fig3}, top panels, what is the
error in the slopes of the fit?}

\discuss{Boeche}{The errors are reported in the text.}

\end{discussion}

\end{document}